\documentstyle[11pt,oldlfont]{article}
\def\Re{\rlap{\rm I}\mkern3mu{\rm R}}
\def\Zed{\rlap{\rm \mbox{\sf Z}}\mkern3mu{\rm \mbox{\sf Z} }}

\def\Ac{{\cal S}}
\def\Lg{{\cal L}}

\def\Hb{{\cal H}}

\def\Bb{{\cal B}}
\def\Mc{{\cal M}}
\def\Kc{{\cal K}}

\def\w{\mbox{\boldmath$\omega$}}

\def\li{\mbox{$\mathcal{L}$}}

\def\bi{\bibitem}

\begin{document}

\thispagestyle{empty}

\begin{flushright} hep-th/0010098 \\
                   AEI-2000-064  \end{flushright}

\vspace*{0.2cm}

\begin{center}{\LARGE {Brane-World charges}} 

\vskip0.2cm

Sebasti{\'a}n Silva

\vskip0.2cm

Max Planck Institut f{\"u}r Gravitationsphysik, Albert Einstein Institut,\\
Am M{\"u}hlenberg 5, D-14476 Golm, Germany\footnote{This work has been
partly supported by the EU TMR contract 
ERBFMRXCT96-0012.
}

\vskip0.2cm

\begin{minipage}{12cm}\footnotesize

{\bf ABSTRACT}
\bigskip

As opposed to usual Einstein gravity in four dimensions, the
Brane-World scenario 
allows the construction of a {\it local} density of gravitational
energy (and also of momentum, of angular momentum, etc\dots ). This is a direct
consequence of the hypothesis that our universe is located at the boundary
of a five-dimensional diffeomorphism invariant manifold.

We compute 
these Brane-World densities of charge using the Lanczos-Israel
boundary conditions. 
 To proceed, we implement an explicitely covariant generalization of the
Hamiltonian procedure of Regge and Teitelboim given in a previous work.

We finally study two simple Brane-World examples.

\bigskip
\end{minipage}
\end{center}
\newpage

\section{Introduction}\label{smintro}

The idea that our four-dimensional universe could be the boundary of
a five-dimensional spacetime has
been recently revived  
as an alternative to Kaluza-Klein compactification \cite{RS} (for
similar suggestions, see also \cite{Go}).
The classical theory in the
bulk is nothing but ordinary Einstein gravity. A Brane-World is 
then located 
at one boundary (say at $r=0$) and satisfies the Lanczos-Israel \cite{LI}
junction conditions. Our results will be independent on what is
located on the other side, say at $r=L$ or $r=\infty$.

As opposed to Kaluza-Klein prescription, gravity is allowed
to propagate in the extra-dimension in the Brane-World
scenarios. Therefore, the effective 
gravitational equations of motion are not the Einstein ones in one
dimension less but the modified version derived in \cite{SMS}. Another
big difference concerns 
the definition of conserved charges
associated with diffeomorphisms, and in particular the definition of energy.
The main purpose of this manuscript is to study this problem in detail.

The ``basic rule'' for conserved charges in gauge invariant theories is the
following: A conserved charge associated with a {\it gauge} symmetry in 
$D$ dimensions  
behaves like a conserved charge associated with a {\it global}
symmetry in $(D-1)$ dimensions. Let us make this more precise.

A global symmetry produces a well defined density of charge in a
$D$-dimensional spacetime through
the Noether construction. This density of charge is in fact given by the
pullback of the corresponding Noether current on a Cauchy
hypersurface $\Sigma_{t}$, namely 
$J^{\mu} t_{\mu}$ (with $t_{\mu }$ the normal to $\Sigma_{t}$).

For a gauge symmetry, this is not anymore
true in general. In fact, the associated Noether current is
generically not gauge
invariant and then not well-defined locally. On the other hand,
a gauge symmetry generates a local density of
charge at {\it each boundary} (denoted by $\Hb_{r}$) of the
$D$-dimensional spacetime. The 
Noether current is then replaced by a two index tensor, called {\it
superpotential} $U^{\mu \nu}=-U^{\nu \mu }$. This tensor is in
general only well-defined (that is covariant) at the boundary $\Hb_{r}$ considered. The density of
charge (at a  
fixed time) is then given by the pullback of this superpotential on
the closed manifold $\Bb_{r}=\Hb_{r}\cap \Sigma _{t}$, namely $U^{\mu
\nu}t_{\mu} n_{\nu}$, with $n_{\mu }$ the normal to $\Hb_{r}$.
The
example of general relativity is well-known: we cannot give a well
defined local 
density of energy in the bulk (the so-called energy-momentum
pseudo-tensors are not covariant) but only at spatial infinity, the
ADM mass.

Suppose now that the bulk spacetime has more boundaries,
for instance an ``isolated horizon'' \cite{ACK} or a Brane-World.
Then, analogously
to the ADM mass at spatial infinity, the bulk diffeomorphism symmetry
allows the construction of a local 
density of mass at {\it each} of these new boundaries in a completely
independent way\footnote{The notion of a local black hole mass computed on
an isolated horizon  is for
instance defined
 in \cite{ACK}.}. We can then talk about  ``holographic charges'' \cite{JS3}.

For concrete calculations of these superpotentials, a
precise knowledge of the boundary conditions is needed.
For the brane-boundary example we are interested in, these are 
the Lanczos-Israel junction conditions. We can then compute
the superpotential associated with the five-dimensional diffeomorphism
invariance on this Brane-World\footnote{As well as an
ADM mass if the five-dimensional spacetime is also asymptotically flat
at $r=\infty$. However this mass cannot be measured since we are not
allowed to escape from the Brane-World.}. 
Following the above ``basic rule'', 
the energy is now a {\it local
density} in our four-dimensional boundary-universe. We compute a
general expression for this density of energy\footnote{\dots and for all the
other charges associated with the diffeomorphism symmetry, as for
instance momentum or
angular momentum.} in any dimension $D$ in section \ref{bwc}. This
is the main result of this manuscript. 

Note finally that some aspects of the Brane-World problem were studied
by relativists in one dimension less. In fact, the situation is
completely analogous to the case of an infinitely thin shell of dust
embedded in a four-dimensional spacetime. Then, the results presented
here could be also useful in this context.

In section \ref{conser}, we recall the construction of conserved charges
associated with gauge symmetries. In section \ref{pureg} we fix the conventions
and the notations for a $D$-dimensional gravity bounded by a
Brane-World. We emphasize on the known relation between the
Lanczos-Israel conditions and the variational principle.
The charges due to the $D$-dimensional diffeomorphism invariance are
computed in section \ref{bwc}. The final result namely equation
(\ref{ghhh}) (together
with (\ref{that})) is then discussed.
The purpose of section \ref{scfiel} is to check if this result
is modified by the introduction of scalar fields in the bulk, with
first Dirichlet (\ref{dbc}) and then Neumann (\ref{nbd}) boundary conditions.
We finish in section \ref{exampl} with two simple Brane-World examples
where our derived 
formula for the energy can be easily compared with the Hamiltonian
result. With both method, we find that the {\it energy vanishes}.

Part of this work is also presented in \cite{SiC}.

\section{Conserved charges in gauge theories}\label{conser}

An Action invariant under one global
symmetry produces one conserved charge. This charge is given by
the integral of the Noether current on a Cauchy
hypersurface $\Sigma _{t}$.

Assume now that the Lagrangian is invariant under a gauge (that is local)
symmetry. Let us also denote by $\Bb_{r}$, $r=\{1,\dots ,n \}$ all the
disconnected components of the boundary of a (partial) Cauchy
hypersurface, namely $\partial \Sigma _{t}=\sum_{r=1}^{n}
\Bb_{r}$. Then 
the conserved charges associated with the gauge symmetry (see
\cite{JS1,Si,SiT,JS3,JS4} and references therein):

\begin{enumerate}
\item [] $\bullet$ can be computed on each $\Bb_{r}$ in a completely
independent way by the formula,
\begin{equation}\label{ccc}
Q_{\xi} = \int_{\Bb_{r}} U^{\mu \nu}_{\xi} d\Sigma_{\mu \nu},
\end{equation}
where $d\Sigma_{\mu \nu }$ is the volume element on $\Bb_{r}$;
\item [] $\bullet$  depend strongly on the boundary conditions imposed
on $\Bb_{r}$.
\end{enumerate}

The density $U^{\mu \nu}_{\xi}$, called superpotential, is
antisymmetric in its upper indices 
and depends explicitely on the gauge parameter $\xi^{a} (x)$.
We now recall the construction of $U^{\mu \nu}_{\xi}$.

The general
``recipe'' goes as follows \cite{Si}: First, let us suppose that the gauge
symmetry transformation
laws of the fields\footnote{The index $^{i}$ labels the set
of all fields (even auxiliary) present in the Lagrangian.}
$\varphi^{i}$ can be written as:

\begin{equation}\label{ggh}
\delta _{\xi} \varphi^{i}= \partial_{\mu} \xi^{a} \Delta^{\mu 
i}_{a} + \xi^{a} \Delta^{i}_{a}.
\end{equation}

We assume that these transformation laws contain no terms
proportional to higher 
derivatives of the local gauge parameter $\xi^{a} (x)$. Naturally,
the index $^{a}$ labels the set of gauge symmetries. The
quantities $\Delta^{\mu i}_{a} (\varphi)$ and
$\Delta^{i}_{a}(\varphi)$ 
given by the gauge symmetry considered are functionals of the
fields (and their first derivatives). 

The next step is then to construct the following tensor

\begin{equation}\label{wdef}
W_{\xi}^{\mu} := \xi^{a} \Delta^{\mu i}_{a} \frac{\delta
\Lg}{\delta \varphi^{i}}
\end{equation}
where we used the definition (\ref{ggh}) and the last term refers to
the equations of motion of $\varphi^{i}$. 

Now comes a very important point: We assume that our theory has been
rewritten in a {\it first order} form. That means, we require that
both, the equations of motion $\frac{\delta \Lg }{\delta \varphi^{i}}$
and the transformation laws $\delta_{\xi} \varphi^{i}$ (\ref{ggh}), 
to depend on the fields and at most their first 
derivatives\footnote{That is, we assume that $\frac{\delta \Lg }{\delta \varphi^{i}}$
and $\delta_{\xi} \varphi^{i}$ {\it do not depend} on $\partial^{2}
\varphi$, $\partial^{3} \varphi$, etc\dots In general, some auxiliary
fields are needed to construct 
these {\it first order} formalisms.}.
We require these restrictions to give rigorous and general proofs
to support the method proposed in \cite{Si}. For more details, see
also \cite{SiT,JS4}.

Then, an {\it arbitrary} variation of the superpotential (which defines the
conserved charge through (\ref{ccc})) should satisfy:

\begin{equation}\label{deful}
\delta U^{\mu \nu}_{\xi}=-\delta \varphi^{i} \frac{\partial
W_{\xi}^{\mu}}{\partial \partial_{\nu} \varphi^{i}}
\end{equation}

The last step is then to ``integrate'' equation (\ref{deful}) {\it using the
boundary conditions} on $\Bb _{r}$. By ``integrate'' we mean: using the 
imposed boundary conditions, we should be able to rewrite the rhs of
(\ref{deful}) as $\delta (\mbox{\it something})$.

The charge (\ref{ccc}), with $U_{\xi}^{\mu \nu}$ satisfying
(\ref{deful}), will be {\it conserved} if the boundary conditions on
$\Bb_{r}$ are compatible with some variational principle \cite{Si,SiT,JS4}.

The simplest examples of Yang-Mills, p-forms or Chern-Simons theories
can be found in \cite{Si, SiT}. The cases of gravity and supergravity
at spatial infinity are treated in \cite{JS3} and \cite{HJS}
respectively. 
The purpose of this manuscript is to study gravity bounded by a
Brane-World.

For another approaches to compute superpotentials that do not
emphasize the boundary conditions see 
\cite{BCJ,BFF,GMS}. For related works and references on conservation laws
in field theories, see also \cite{AT}.

\section{Pure gravity bounded by a Brane-World}\label{pureg}

We consider a $(D-2)$-brane located at
the boundary of a
D-dimensional spacetime.
In vacuum, with signature mostly plus, we start with the following
action\footnote{We use the convention $4\kappa ^{2}=16 \pi G$.}:

\begin{equation}\label{lagacf}
\Ac = \int_{\Mc} \Lg:= \int_{\Mc} \left( \frac{\sqrt{\left|
g\right|}}{4 \kappa ^{2}} R - \sqrt{\left|
g\right|} \Lambda - \partial_{\mu} S^{\mu} - \partial_{\mu} \left(
\frac{n^{\mu}}{N^{2}}
\Lg_{bra} \delta (\chi)\right)\right)
\end{equation}
where

\begin{equation}\label{sdef}
S^{\mu}:=\frac{\sqrt{\left|
g\right|}}{4 \kappa ^{2}} ( \Gamma ^{\mu }_{\rho \sigma} g^{\rho
\sigma}- \Gamma ^{\sigma }_{\rho \sigma } g^{\rho\mu }).
\end{equation}

The Lagrangian (\ref{lagacf}) is pure gravity in $D$ dimensions, with some
special boundary terms which take into account the presence of
the brane. We discuss this point in detail below.

 The brane is a $(D-1)$-dimensional 
hypersurface (which is assumed timelike) embedded in the
$D$-dimensional spacetime through the
constraint:

\begin{equation}\label{cone}
\chi (x) =0,
\end{equation}
for some given bulk function $\chi$.

The (spacelike) {\it in-going} vector normal to this brane is given by:

\begin{equation}\label{nove}
n_{\mu}:=\partial_{\mu }\chi, \ \  N:=\sqrt{g^{\mu \nu
}n_{\mu } n_{\nu} }, \ \  \hat{n}_{\mu}:=\frac{n_{\mu } }{N}. 
\end{equation}

Therefore, the Stokes theorem gives

\begin{equation}\label{integral}
\int_{\Mc } \partial_{\mu } \left(\Sigma ^{\mu }  \delta (\chi)
\right) = - \int_{brane} n_{\mu} \Sigma ^{\mu }.
\end{equation}
In particular, the last term of equation (\ref{lagacf}) is nothing but
$\int_{brane} \Lg_{bra}$. In the following, we will also assume that
$\Lg_{bra}$ depends on the induced metric $h_{\mu \nu }:=g_{\mu \nu
}-\hat{n}_{\mu }\hat{n}_{\nu }$ and eventually on some brane-fields.

For simplicity, we choose 
the transverse $D^{th}$ coordinate $r$ such that

\begin{equation}\label{specoo}
r=\chi (x).
\end{equation}

The brane is then located at $r=0$ and $n_{\mu}=\delta_{\mu }^{r}$. The transverse direction can be
either a usual interval $r\in [0,L]$ or a semi-infinite one, $r\in
[0,\infty)$. The orbifold case, namely $r\in S^{1}/\Zed_{2}$, is up to
a factor of two, analogous to the ordinary interval
case. This can be 
taken into account by the replacements
$\Lg_{bra}\rightarrow \Lg_{bra}/2$ (and then $T^{bra}_{\rho \sigma}
\rightarrow T^{bra}_{\rho \sigma}/2$) in the following formulae, and in
particular in the main result (\ref{ghhh}). We explain this point in
detail in the Appendix.

We work with a first order formalism (namely the Palatini one)
where the metric $g^{\mu \nu }$ and the connection
$\Gamma^{\rho}_{\mu \nu}$ are assumed to be {\it independent
fields}. Then, the scalar curvature in (\ref{lagacf}) is a
functional of the  metric, of the connection and its first derivatives,
namely, $R=R (g, \Gamma, \partial \Gamma)$. Note that $\Lg =\Lg (g,
\partial g, \Gamma)$ of (\ref{lagacf}) is the so-called Einstein
Lagrangian which does not depend on $\partial \Gamma $ and
differs from the Hilbert one (namely the scalar curvature) by the
surface term (\ref{sdef}). 

The Euler-Lagrange variation of a total derivative vanishes. The
surface terms in (\ref{lagacf}) then do not change the bulk equations
of motion, which are given by:

\begin{eqnarray}
\frac{\delta \Lg }{\delta g^{\mu \nu}} &=&  \frac{\sqrt{\left|
g\right|}}{4 \kappa^{2}} G_{\mu \nu} +\frac{\sqrt{\left|
g\right|}}{2} g_{\mu \nu } \Lambda \label{equati1} \\ 
\frac{\delta \Lg }{\delta \Gamma ^{\rho}_{\mu \nu}}  &=& - \frac{1}{4
\kappa^{2}} \nabla_{\sigma} \left(\sqrt{\left|
g\right|} g^{\mu \nu } \delta^{\sigma}_{\rho} - \sqrt{\left|
g\right|} g^{\sigma (\nu} \delta^{\mu) }_{\rho}\right). \label{equati2}
\end{eqnarray}
where the Einstein tensor is as usual $G_{\mu \nu}:=R_{\mu \nu
}-\frac{1}{2}g_{\mu \nu } R$.
Note that for $D\geq 3$, the equation (\ref{equati2}) is equivalent to
$\nabla_{\rho}g^{\mu \nu}=0$ which defines the connection in term of
the metric and its first derivatives.

The surface term\footnote{Together with equation
(\ref{nsimpl}), this surface term (\ref{sdef}) is nothing but the
Gibbons and Hawking \cite{GH} extrinsic curvature boundary term. In
fact, using these equations we can check that $-n_{\mu
}S^{\mu }=\sqrt{\left| h \right|}/ (2\kappa ^{2}) {\cal K}$, with
${\cal 
K}= h^{\mu \nu } \nabla_{\mu }\hat{n}_{\nu }$ the trace of the extrinsic
curvature.} $\partial_{\mu } S^{\mu}$ in (\ref{lagacf}) is needed
in order to  
satisfy the variational principle and
the junction 
conditions on the brane. 
In fact $\delta \Ac   =0 \Leftrightarrow 
(\ref{equati1})= (\ref{equati2})=0$ is only true if the
remaining boundary term 
vanishes. For the precise action (\ref{lagacf}) (and recalling equation
(\ref{integral})), this condition becomes:

\begin{equation}\label{bdcond}
\int_{brane}\left( - n_{\mu}  \frac{\partial
\Lg}{\partial \partial_{\mu} g^{\rho \sigma}} \delta g^{\rho
\sigma} + \delta (\Lg_{bra}) \right) = 0 
\end{equation}
with
\begin{equation} \label{qdef00}
\frac{\partial
\Lg}{\partial \partial_{\mu} 
g^{\rho \sigma}}\delta g^{\rho \sigma} = - \frac{1}{4 \kappa ^{2}} \left(
\Gamma ^{\mu }_{\rho 
\sigma} \delta (\sqrt{\left| g\right|} g^{\rho \sigma}) -  \Gamma
^{\sigma }_{\rho \sigma } \delta (\sqrt{\left| g\right|} g^{\mu \rho})
\right) =: K^{\mu}_{\rho \sigma}
\delta g^{\rho \sigma},
\end{equation}
and therefore,
\begin{equation}\label{qdef}
K^{\mu}_{\rho \sigma} = - \frac{\sqrt{\left|
g\right|}}{4 \kappa^{2}} \left(\Gamma^{\mu}_{\rho \sigma} -\frac{1}{2}
g_{\rho \sigma} \Gamma^{\mu}_{\alpha \beta} g^{\alpha \beta}-\Gamma
_{\alpha (\rho }^{\alpha}\delta^{\mu }_{\sigma)} +\frac{1}{2}
\Gamma^{\alpha }_{\beta \alpha} g^{\beta \mu } g_{\rho \sigma}
\right).
\end{equation}

Note that the equations (\ref{sdef}) and (\ref{qdef00}), (\ref{qdef})
imply the identities (to be used in the following section \ref{bwc}):  

\begin{eqnarray}
 K^{\mu}_{\rho \sigma} g^{\rho \sigma } &=& \frac{D-2}{2} S^{\mu},
\label{idenu} \\
 K^{\mu}_{\rho \sigma} \delta g^{\rho \sigma } &=& -\delta S^{\mu } +
\frac{\sqrt{\left| g\right|}}{4 \kappa^{2}}g^{\rho \sigma
}\left(\delta \Gamma ^{\mu }_{\rho \sigma }-\delta \Gamma ^{\alpha }_{\alpha
\rho }\delta ^{\mu }_{\sigma } \right). \label{idenu2}
\end{eqnarray}

Let us now assume that the brane-matter
Lagrangian $\Lg_{bra}$ depends on the pulled-back metric  $\left. (h_{\mu
\nu}=g_{\mu \nu }-\hat{n}_{\mu }\hat {n}_{\nu }) \right|_{brane}$ (and possibly
on other fields) but not on the
connection. 
The condition (\ref{bdcond}) together with the definition
(\ref{qdef00}) then implies two equations:

\begin{eqnarray}
\int_{brane} \delta g^{\rho \sigma} \left( -n_{\mu} K^{\mu}_{\rho
\sigma} - \frac{\sqrt{\left| g\right|}}{2} T_{\rho \sigma }^{bra} \right) &=&
0 \label{junc} \\ 
\int_{brane} \left. \delta \Lg_{bra}\right|_{\delta g=0} &=& 0 \label{eqmbr}
\end{eqnarray}
where 
\begin{equation}\label{enmten}
T_{\rho \sigma }^{bra} := - \frac{2}{\sqrt{\left|
g\right|}} \frac{\delta \Lg_{bra}}{\delta g^{\rho \sigma}} 
\end{equation}
denotes the brane-matter energy-momentum tensor which satisfies $T_{\rho
\sigma }^{bra}n^{\sigma}=0$.

As we recall in the Appendix,
the boundary equation (\ref{junc}) is equivalent to
the Lanczos-Israel \cite{LI} junction condition: 

\begin{equation}\label{extrc0}
-\frac{1}{2 \kappa^{2}} \Kc_{\mu \nu} = T_{\mu
\nu}^{bra}-\frac{1}{D-2}h_{\mu \nu} T^{bra} 
\end{equation}
with $\Kc_{\mu \nu}:= h^{\rho}_{\ (\mu} h^{\sigma}_{\ \nu)}
\nabla_{\rho} \hat {n}_{\sigma}$ the extrinsic curvature on
the brane, $ T^{bra}:=h^{\rho \sigma } T_{\rho
\sigma}^{bra}$ and $2\kappa ^{2}=8 \pi G$.

The relation between the variational principle (together with the boundary term
(\ref{sdef})) and the junction condition (\ref{extrc0}) 
was first realized by
Hayward and Louko \cite{HL} using the Hamiltonian formalism, for a
shell of dust in four-dimensional gravity (for more recent work, see
\cite{Gl}
 and also \cite{CR,DM,BV} in the Brane-World context).

On the other hand, the equation  (\ref{eqmbr}) generates the equations
of motion of the Brane-World fields.

\section{The Brane-World charges}\label{bwc}

We assumed that the D-dimensional manifold is bounded by one
brane. Following section \ref{conser}, we
can therefore 
compute the superpotential associated with the
($D$-dimensional) diffeomorphism invariance, at the boundary where this
Brane-World is located.

With the Brane-World scenario in mind, we 
also assume that the D-dimensional metric is

\begin{equation}\label{usform}
ds^{2} = e^{2 A (r)}h_{ij} (x) dx^{i} dx^{j} + dr^{2},
\end{equation}
with the label $i$ running from $0$ to $(D-2)$ and $h_{ij} (x)$ some
``World'' metric.

Then, with the Brane-World ans{\"a}tze (\ref{specoo}) and (\ref{usform}), the
normal vector (\ref{nove}) takes the simple form:

\begin{equation}\label{nsimpl}
n_{\mu }=\hat{n}_{\mu} = \delta _{\mu}^{r}, \ \ n^{\mu }=\hat{n}^{\mu}
= \delta^{\mu r}, \ \ N=1.
\end{equation}

The diffeomorphism symmetry parameter $\xi^{\mu }$ is not arbitrary but has
to be compatible with the boundary conditions: 

\begin{enumerate}
\item [] $\bullet$ First, it should
leave the constraint (\ref{cone}) unchanged:

\begin{equation}\label{unj}
\li_{\xi} \chi =\xi ^{\mu }\partial_{\mu }\chi = 0 \ \ \ \mbox{\it on
the brane},
\end{equation}
with $\li_{\xi}$ the Lie derivative.
Note that equation (\ref{unj}) together with (\ref{nove}) implies:

\begin{equation}\label{tang} 
\xi^{\mu} n_{\mu} = 0, \ \ \li_{\xi} \hat{n}_{\mu } = -\hat{n}_{\mu } \xi^{\rho}
\partial_{\rho } (\ln N) \ \ \ \mbox{\it on
the brane}.
\end{equation}

Therefore, the diffeomorphism parameter $\xi^{\mu }$ should be tangent
to the brane.
For our simple ans{\"a}tze (\ref{usform}) and (\ref{nsimpl}), this becomes: 

\begin{equation}\label{jjj}
\xi^{r}=0, \ \ \li_{\xi} n_{\mu} = 0  \ \ \ \mbox{\it on
the brane}.
\end{equation}

\item [] $\bullet$ Second, the boundary condition (\ref{junc}) should be satisfied by the
transformed fields $\tilde{\varphi}^{i}=\varphi^{i}+\li_{\xi } \varphi^{i}$,
where $\varphi^{i}$ goes for all the fields of our theory. This
condition leads to:

\begin{equation}\label{condl}
\left. \xi ^{\nu } \partial_{\nu }\left(n_{\mu} K^{\mu }_{\rho \sigma }
-T^{bra}_{\rho \sigma}\right)\right|_{brane} =0.
\end{equation}

However, $\xi^{\mu}$ is tangent to the brane by equation
(\ref{tang}). Therefore, the last condition (\ref{condl}) is automatically
satisfied from the junction condition (\ref{junc}).

\item [] $\bullet$ Third, the ansatz (\ref{usform}) should  be
unchanged. This, together with (\ref{jjj}), implies:
\begin{equation}\label{notmod}
\li_{\xi}g_{\mu r} = 0 \Rightarrow -\partial_{r}\xi^{\mu} = \li_{\xi}
n^{\mu} = 0.
\end{equation}

This last condition could be relaxed in more general cases where
the metric is not required to be of the form (\ref{usform}).

\end{enumerate}
In the following, we
assume that $\xi^{\mu }$ satisfies both (\ref{jjj}) and
(\ref{notmod}).

The general variation of the superpotential associated with
diffeomorphisms,
for any $D$-dimensional pure gravity and for any boundary condition was
given in \cite{JS3} in differential forms (where calculations are
easier). We translate here in components the steps of this
calculation.
There is however an important point: we cannot use in a
straightforward way the method summarized in section \ref{conser}. In
fact, in the Palatini formalism, the transformation law 
(under a diffeomorphism) of the connection contains {\it second}
derivatives of the gauge parameter, namely $\delta_{\xi} \Gamma
=\partial^{2} \xi+\mbox{\it other}$. Therefore, it is not of the form
(\ref{ggh}). This technical problem is cured using the
so-called Affine-$GL (D,\Re )$ gravity \cite{JS1,JS3} (see also the
discussion in \cite{SiC}).
Then, we are not going to re-compute but {\it translate} to components
the results of \cite{JS3} for illustrative purposes.

In particular, the $W^{\mu}_{\xi}$ tensor of
(\ref{wdef}) is given by\footnote{The fact that the tensor
(\ref{wten}) contains first derivatives of the gauge parameter
$\xi^{\mu}$ while it does not in the definition (\ref{wdef}) is not a
contradiction. As we have mentioned, formula (\ref{wten}) cannot be used in the
Palatini formalism due to the presence of second derivatives of the gauge
parameter in the transformation law of the connection. The rigorous
way to proceed is then to use the Affine-$GL (D,\Re)$ formalism \cite{JS1,JS3}
which gives the result (\ref{wten}).}:

\begin{equation}\label{wten}
W^{\mu}_{\xi}=-2 \xi^{\rho} g^{\sigma \mu} \frac{\delta
\Lg}{\delta g^{\rho \sigma}}+ \nabla _{\rho} \xi^{\sigma}
\frac{\delta\Lg}{\delta \Gamma ^{\sigma }_{\mu \rho}}.
\end{equation}

A very similar expression exists for the vielbein formalism (in its
first order form), namely:

\begin{equation}\label{wtenv}
W^{\mu}_{\xi}= \xi^{a} \frac{\delta
\Lg}{\delta e^{a}_{\mu}}+ D_{b} \xi^{a}
\frac{\delta \Lg}{\delta \w ^{a }_{\mu b}}
\end{equation}
where the internal Lorentz indices are denoted by Latin letters, and
$\xi^{a}:=\xi^{\rho} e^{a}_{\rho}$ and $D_{b} \xi^{a}:=e_{b}^{\rho}
(\partial_{\rho} \xi^{a} +\w_{\rho c}^{a}\xi^{c})$.

The similarity between (\ref{wten}) and (\ref{wtenv}) can be explained
by their common origin in the Affine-$GL (D,\Re)$ gravity.
Again, following the results of the work \cite{JS3}, 
 the
variation of the superpotential  due
to the diffeomorphism invariance 
of the gravitational theory is given by,


\begin{eqnarray}
\delta U_{\xi}^{\mu \nu }
&=& \frac{1}{4 \kappa ^{2}} \left( 2\  \nabla_{\sigma } \xi^{\tau}
\delta \left( \sqrt{\left| 
g\right|} g^{\sigma [\mu }  \delta^{\nu] }_{\tau
} \right) + 6\  \delta \left( \Gamma^{\tau }_{\rho \sigma}\right)
\sqrt{\left| 
g\right|} g^{\sigma [\mu}  \delta^{\nu}_{\tau } \xi
^{\rho ]}  \right) \ \ \label{encom}\\
 &=&  \frac{1}{4 \kappa ^{2}} \left( 2\  D^{a} \xi^{b} \delta \left(  \left|
e\right|  e^{[\mu}_{a} e^{\nu] }_{b} \right)  + 6\   \delta \left( \omega^{ab}_{\rho}\right)  \left| e \right|
e^{[\mu}_{a} e^{\nu}_{b} \xi
^{\rho ]}
   \right)\label{enorth}
\end{eqnarray}
in the Palatini and vielbein formalisms respectively.

The equations (\ref{encom}) and (\ref{enorth}) are valid when pulled-back on {\it any}
$\Bb_{r}$ (and for {\it any} boundary condition compatible with the
variational principle, see section
\ref{conser}). In particular, they were used in \cite{JS3} to derive
the KBL superpotential \cite{KBL} at spatial infinity.

Now, assuming $\delta \xi^{\rho}=0$ and using the identity
(\ref{idenu2}), the equation (\ref{encom}) can be
rearranged as\footnote{Schematically, the calculation from equation
(\ref{encom}) to equation (\ref{begeq}) goes as follows: the first term of
equation (\ref{encom}) plus two of the six $\delta \Gamma $'s  gives
 $\delta (\ ^{\mbox{\tiny \it Ko}}U_{\xi}^{\mu \nu })$. The four
remaining $\delta \Gamma $'s give 
$(\delta g^{\rho \sigma } K^{\mu }_{\rho \sigma }+ \delta S^{\mu })
\xi ^{\nu }-\mu \leftrightarrow \nu $ using the identity
(\ref{idenu2}).}:

\begin{equation}\label{begeq}
 \delta U_{\xi}^{\mu \nu } = \delta \left(\ ^{\mbox{\tiny \it
Ko}}U_{\xi}^{\mu \nu } + S^{\mu} \xi^{\nu}-S^{\nu}
\xi^{\mu} \right)+\delta g^{\rho \sigma } K_{\rho \sigma }^{\mu}
\xi^{\nu}- \delta g^{\rho \sigma } K_{\rho \sigma }^{\nu} \xi^{\mu}
\end{equation}
where we have defined

\begin{equation}\label{usdef}
\ ^{\mbox{\tiny \it Ko}}U_{\xi}^{\mu \nu }:=\frac{\sqrt{\left|
g\right|}}{4 \kappa ^{2}} (\nabla^{\mu} \xi^{\nu}-\nabla^{\nu}
\xi^{\mu}),
\end{equation}
for the (1/2) Komar superpotential \cite{Ko}.

For the Brane-World boundary, we 
contract equation (\ref{begeq})
along $t_{\mu}$ and $n_{\nu}$, with $t_{\mu}$ some
timelike co-vector (tangent to the brane).
With the condition (\ref{tang}), the equation
(\ref{begeq}) {\it on the brane} becomes:

\begin{eqnarray}
\delta \left(  U_{\xi}^{\mu \nu }\  t_{\mu} n_{\nu}\right)&=&\delta \left(
 \ ^{\mbox{\tiny \it
Ko}}U_{\xi}^{\mu \nu }\  t_{\mu} n_{\nu} -n_{\nu} S^{\nu}\ 
t_{\mu} \xi^{\mu} \right) - \delta g^{\rho \sigma } n_{\nu}K_{\rho
\sigma }^{\nu} \  t_{\mu}\xi^{\mu} \nonumber \\
&=& \delta \left( \  ^{\mbox{\tiny \it
Ko}}U_{\xi}^{\mu \nu }\ 
t_{\mu} n_{\nu} +\frac{\sqrt{\left|
g\right|}}{D-2} T^{bra}\ 
t_{\mu} \xi^{\mu}  -  \Lg_{bra}\  t_{\mu}\xi^{\mu}\right)\label{intb}
\end{eqnarray}
where we used equations (\ref{bdcond}), (\ref{qdef00}), (\ref{idenu}) and
(\ref{junc}).

Therefore, the superpotential is (up to some global constant): 

\begin{equation}\label{grama}
U_{\xi}^{\mu \nu }\ t_{\mu} n_{\nu}  = \  ^{\mbox{\tiny \it
Ko}}U_{\xi}^{\mu \nu }\  t_{\mu} n_{\nu} + \left( \frac{\sqrt{\left|
g\right|}}{D-2} T^{bra} 
- \Lg_{bra} \right) \ 
t_{\mu}\xi^{\mu} + C^{t}.
\end{equation}

The above number $C^{t}$  can be fixed by imposing the vanishing
of the superpotential for some reference solution. It just defines the zero
point energy. We will set $C^{t}=0$ in the following.

The expression (\ref{grama}) can be simplified. If fact,
using again 
the conditions (\ref{jjj}) and (\ref{notmod}), the (1/2) Komar superpotential (\ref{usdef}) along $t_{\mu}$
and $n_{\nu}$ can be rewritten as\footnote{If we relax the constraint
(\ref{notmod}), that is, we do not require the ansatz (\ref{usform}),
a new term appears in the rhs of equation (\ref{komre}), namely $\sqrt{\left|
g\right|}/ (4\kappa ^{2})t_{\mu} \li_{\xi} n^{\mu }$. This could be
relevant for more general situations.}

\begin{equation}\label{komre}
\ ^{\mbox{\tiny \it Ko}}U_{\xi}^{\mu \nu } t_{\mu } n_{\nu }:=-\frac{\sqrt{\left|
g\right|}}{2 \kappa ^{2}}  t^{\mu} \xi^{\nu } \Kc_{\mu \nu },
\end{equation}
with $\Kc_{\mu \nu }$ the extrinsic curvature on the brane.

Finally, using once again the boundary conditions (\ref{extrc0}), the
{\it total} 
superpotential (\ref{grama}) simplifies considerably:

\begin{eqnarray}
U_{\xi}^{\mu \nu }\ t_{\mu} n_{\nu}  &=& \sqrt{\left|
g\right|} t^{\mu  } \xi ^{\nu } T_{\mu
\nu}^{bra} - \Lg_{bra}  t_{\mu}\xi^{\mu} \nonumber \\
&=& \sqrt{\left| g\right|} t^{\mu  } \xi ^{\nu } \hat{T}_{\mu
\nu}^{bra}, \label{ghhh}
\end{eqnarray}
where we have defined:

\begin{equation}\label{that}
\hat{T}_{\mu \nu}^{bra}:= -2 \frac{\delta
}{\delta g^{\mu \nu}} \left(\frac{\Lg_{bra}}{\sqrt{\left| g\right|}} \right).
\end{equation}

The quantity (\ref{ghhh}) gives the density of charge at a fixed
time depending on the
vector $\xi^{\mu}$ used (for instance mass,
momentum or angular momentum). The expression
(\ref{that}) is ``almost'' the energy momentum tensor on the
brane (compare with (\ref{enmten})). However, the differences are quite important: 

\begin{enumerate}
\item [] $\bullet$ The tensor (\ref{that}) will {\it vanish} for any
Brane-Lagrangian which depends on the metric only through an overall 
$\sqrt{\left| g\right|}$ factor. See for instance the examples of
section \ref{exampl}.
\item [] $\bullet$ The expression (\ref{that}) gives the total energy
(and other 
charges as linear and angular momentum), including the {\it
gravitational contribution}. In the
Brane-World scenario, the (gravitational) energy is then a {\it
local density}. The total
conserved charge is simply the integral of (\ref{ghhh}) on a
$(D-2)$-dimensional Cauchy hypersurface on the brane at fixed time
(namely, on $\Bb_{0}$ in the notation of section \ref{conser}).
\item [] $\bullet$ The positivity of energy would require some modified 
{\it energy condition}, namely $\hat{T}_{\mu \nu}^{bra}t^{\mu}\xi^{\nu}
\geq 0$, with $\xi^{\mu}$ some timelike vector.
\end{enumerate}

Note also that the Nester superpotential \cite{Ne} was used in
previous works \cite{Cv} in order to define the gravitational energy
on a Brane-World (Domain Wall). However, this
superpotential\footnote{The Nester superpotential is nothing but a
covariant expression for the ADM mass. Moreover, it appears naturally
in any supergravity in the commutation of two supercharges (also
defined at {\it spatial infinity}), see for instance
\cite{HJS} and references therein.} is
associated with {\it asymptotically flat} boundary conditions and not
with the Lanczos-Israel ones (\ref{extrc0}).

It would also be interesting to compare our result (\ref{that}) with
the effective energy-momentum stress tensor derived (at the linearized
level) in \cite{GK}\footnote{I thank S.B. Giddings for pointing out
this suggestion.}.

Finally, it should be possible to recover the result (\ref{ghhh})
using the Hamiltonian formalism together with the Regge and Teitelboim
prescription \cite{RT}.

\section{The bulk scalar fields}\label{scfiel}

Let us now introduce some scalar
fields $\phi^{I}$ in the bulk.
The corresponding action is

\begin{equation}\label{newgr}
\Lg_{sc}:= \sqrt{\left| g\right|} \left( -\Pi^{\mu}_{I} 
 \partial_{\mu}
\phi^{I} +\frac{1}{2} g_{\mu \nu }\  G^{IJ} (\phi)
\Pi^{\mu}_{I}\Pi^{\nu}_{J} - V (\phi)   \right),
\end{equation}
with $G_{IJ} (\phi)$ some given moduli space metric. Remember that
we are using exclusively first order formalisms; that is why we
introduced the auxiliary fields $\Pi^{\mu}_{I}$.

The equations of motion derived from (\ref{newgr}) are:

\begin{eqnarray}\label{nveq}
\frac{\delta \Lg_{sc}}{\delta \phi^{I}} &=& \partial_{\mu}\left(\sqrt{\left|
g\right|} \Pi^{\mu}_{I} \right) +\frac{\sqrt{\left|
g\right|}}{2} g_{\mu \nu } G^{JK}_{,I} \Pi^{\mu}_{J}\Pi^{\nu}_{K}- \sqrt{\left|
g\right|} V_{,I}\label{eqmop} \\
\frac{\delta \Lg_{sc}}{\delta \Pi^{\mu}_{I}} &=& \sqrt{\left|
g\right|} \left( g_{\mu \nu } G^{IJ}
\Pi_{J}^{\nu}- \partial_{\mu}\phi^{I}  \right)  \label{auxeq} \\
\frac{\delta \Lg_{sc}}{\delta g^{\mu \nu}} &=& -\frac{\sqrt{\left|
g\right|}}{2} \Pi_{I\mu}
G^{IJ} \Pi_{J\nu}-\frac{1}{2}g_{\mu \nu } \Lg_{sc}. \label{eismatt}
\end{eqnarray}

We will study two kind of boundary conditions for the scalar
fields:
\begin{enumerate}
\item [] 1. The Dirichlet condition together with the assumption that
the brane Lagrangian 
$\Lg_{bra}$ (of equation (\ref{lagacf})) depends {\it neither} on the
scalar fields 
$\phi^{I}$ {\it nor} on the auxiliary ones $\Pi ^{\mu }_{I}$.
\item [] 2. The Neumann condition together with the assumption that 
$\Lg_{bra}$ can depend on the scalar fields $\phi^{I}$ (but {\it not}
on $\Pi^{\mu }_{I}$).
\end{enumerate}

\subsection{The Dirichlet boundary conditions}\label{dbc}

We would like to impose some Dirichlet boundary conditions on the
scalar fields, that is, $\phi^{I}=\phi^{I}_{0}=${\it constant} on the brane.
This can be implemented at the variational principle level by adding
the following surface term to the action (\ref{newgr})

\begin{equation}\label{suterm}
  \partial_{\mu}\left(  \sqrt{\left| g\right|} \Pi^{\mu}_{I} (\phi^{I}-\phi^{I}_{0})\right),
\end{equation}
which of course does not modify the equations of motion
(\ref{eqmop}), (\ref{auxeq}) and (\ref{eismatt}).

The variational principle, namely  $\delta\int
\left((\ref{newgr})+ (\ref{suterm}) \right) =0\Leftrightarrow
(\ref{eqmop})= (\ref{auxeq})= (\ref{eismatt}) =0$, is satisfied 
if and only if 

\begin{equation}\label{bdconsc}
-\int_{\mbox{\it \footnotesize brane}}\delta \left( \sqrt{\left| g\right|} n_{\mu}\Pi^{\mu}_{I}\right) (\phi^{I}-\phi^{I}_{0})=0
\end{equation}
which are our imposed Dirichlet boundary conditions.

The purpose is now to prove that 
the result (\ref{ghhh}) is not modified by the
presence of the scalar Lagrangian (\ref{newgr})+(\ref{suterm}).

 Under one diffeomorphism, the scalar fields transform as:

\begin{eqnarray}
\delta _{\xi} \phi^{I} &=& \xi ^{\rho }\partial_{\rho} \phi^{I}\label{tranph}\\
\delta _{\xi} \Pi^{I}_{\mu } &=&  \xi ^{\rho }\partial_{\rho}
\Pi^{I}_{\mu }+\partial_{\mu}\xi ^{\rho }  \Pi^{I}_{\rho  }\label{ptran}
\end{eqnarray}

Then, the tensor $W^{\mu }_{\xi }$ of (\ref{wten}) receives a new
contribution (see definition (\ref{wdef})), namely:

\begin{equation}\label{neww}
\ _{sc}W_{\xi}^{\mu}= -2 \xi^{\rho} g^{\sigma \mu} \frac{\delta
\Lg_{sc}}{\delta g^{\rho\sigma}} - \Pi^{\mu}_{I}
\xi^{\rho}\frac{\delta \Lg_{sc}}{\delta \Pi^{\rho}_{I}}.
\end{equation}

Following equation (\ref{deful}) the variation of the superpotential
(\ref{encom}) receives a ``scalar fields'' contribution:

\begin{eqnarray}
\delta \ _{sc}U^{\mu \nu}_{\xi} &=&  2  \xi^{\rho} g^{\sigma \mu}
\left(\frac{1}{2} g_{\rho \sigma }  \sqrt{\left| g\right|}
\Pi_{I}^{\nu}\right)\delta \phi^{I}-\Pi^{\mu}_{I}
\xi^{\nu} \sqrt{\left| g\right|} \delta \phi^{I} \nonumber\\
&=& \left( \xi^{\mu}\Pi^{\nu}_{I} - \xi^{\nu} \Pi^{\mu}_{I} \right)
\sqrt{\left| g\right|} \delta \phi^{I}. \label{unew}
\end{eqnarray}

Again, the result (\ref{unew}) is valid for {\it any} boundary
condition and then will also be used in the next subsection for
Neumann boundary conditions. Note that the manifest antisymmetry in
$\mu $ and $\nu $ was guaranteed by a theorem \cite{Si}.

As in the previous section \ref{bwc}, we pullback equation (\ref{unew}) on the
brane and use the boundary conditions (\ref{bdconsc}) (and the
condition (\ref{tang})) in order to ``integrate'' it:

\begin{eqnarray}
\delta \left( \ _{sc}U^{\mu \nu}_{\xi} \ t_{\mu} n_{\nu} \right) &=&
 \sqrt{\left| g\right|} n_{\nu}\Pi^{\nu}_{I} \  t_{\mu} \xi^{\mu}
\delta \phi^{I} \nonumber\\
&=& \delta \left(  \sqrt{\left| g\right|} n_{\nu}\Pi^{\nu}_{I} \
t_{\mu} \xi^{\mu} \left( \phi^{I}-\phi^{I}_{0}\right)\right)=0.\label{besca}
\end{eqnarray}

Then, the superpotential (\ref{ghhh}) is not modified by the
scalar fields with Dirichlet boundary conditions (\ref{bdconsc}). Note
that the same is  
true for asymptotically flat spacetimes: The formula for the ADM mass
at spatial 
infinity remains unchanged if some scalar fields with Dirichlet boundary
conditions are included in the bulk.

\subsection{The Neumann boundary conditions}\label{nbd}

Let us turn now to the case of Neumann boundary conditions on the
scalar fields. We first add to the Lagrangian (\ref{newgr}) the
following surface term:

\begin{equation}\label{newbd}
\partial_{\mu }\left(\Pi _{0I}^{\mu} \phi^{I} \right)
\end{equation}
with $\Pi _{0I}^{\mu}$ some given constant (possibly zero).

We now assume that $\Lg_{bra}=\Lg_{bra} (\phi)$. This introduces a
coupling with our pure gravity model of section \ref{bwc}.
The variational principle associated with
(\ref{lagacf})+(\ref{newgr})+(\ref{newbd}) is satisfied if and only if the
following condition on the boundary (brane) holds:

\begin{equation}\label{bdneu}
\int_{brane}\left( -n_{\mu} K^{\mu}_{\rho \sigma} \delta g^{\rho
\sigma}+ \delta (\Lg_{bra}) +\sqrt{\left| g\right|} n_{\mu}\Pi
^{\mu }_{I} \delta \phi^{I} - \delta (n_{\mu} \Pi _{0I}^{\mu}
\phi^{I})\right) = 0.
\end{equation}

The equation (\ref{bdneu}) implies then together with
(\ref{junc}-\ref{eqmbr}) another ``junction equation'' for the scalar
fields on the brane:

\begin{equation}\label{nerrr}
 \sqrt{\left| g\right|} n_{\mu}\Pi
^{\mu }_{I}-n_{\mu}\Pi _{0I}^{\mu}=-\frac{\partial
\Lg_{bra}}{\partial\phi^{I}}.
\end{equation}

This is the usual 
Neumann boundary condition which fixes the normal derivative of the
scalar fields. In many case the constant $\Pi _{0I}^{\mu}$ is set to
zero. We will however consider the general case. 

Now, the variation of the total superpotential is just the sum of
(\ref{begeq}) and (\ref{unew}). When pulled-back on the brane, this
equation gives: 

\begin{eqnarray}
\delta \left(\   _{tot}U_{\xi}^{\mu \nu }\  t_{\mu} n_{\nu}\right)&=&\delta \left(
 \ ^{\mbox{\tiny \it
Ko}}U_{\xi}^{\mu \nu }\  t_{\mu} n_{\nu} -n_{\nu} S^{\nu}\ 
t_{\mu} \xi^{\mu} \right) - \delta g^{\rho \sigma } n_{\nu}K_{\rho
\sigma }^{\nu} \  t_{\mu}\xi^{\mu} \nonumber \\
& & + \sqrt{\left| g\right|} n_{\nu} \Pi^{\nu}_{I} \  t_{\mu} \xi^{\mu}
\delta \phi^{I}. \label{intb2}
\end{eqnarray}

Using the boundary condition (\ref{bdneu}), we find that:

\begin{equation}\label{utot}
 _{tot}U_{\xi}^{\mu \nu }\  t_{\mu} n_{\nu} =  U_{\xi}^{\mu \nu }\
t_{\mu} n_{\nu} +n_{\nu} \Pi^{\nu}_{0I} \phi^{I}\
t_{\mu} \xi^{\mu}
\end{equation}
with $U_{\xi}^{\mu \nu }$ given by (\ref{ghhh}).
Then, the total superpotential is modified only if
$\Pi^{\nu}_{0I}$ is non-zero.

\section{Applications}\label{exampl}

The purpose of this last section is to illustrate the formula (\ref{ghhh})
with some Brane-World examples. For concreteness, we will
start with the following five-dimensional metric (the extension to
higher dimensions is straightforward):

\begin{equation}\label{m5d}
ds^{2}=e ^{2 A (r)} \eta_{ij} dx^{i} dx^{j} + dr^{2},
\end{equation}
with $\eta_{ij}=\{-,+,+,+\}$.

We mostly follow the conventions of \cite{DFGK} (the signature of the
metric is,
however, inverted). We then fix
$\kappa ^{2}=1$ and $\Lambda =0$ in the Lagrangian (\ref{lagacf}). 

We also include one bulk scalar field $\phi (r)$, with Neumann boundary
conditions (\ref{nerrr}) together with $\Pi_{0I}^{\mu}=0$ (see subsection
\ref{nbd}). Finally,
we choose
 
\begin{equation}\label{branla}
\Lg_{bra}=-\frac{\sqrt{\left| g\right|}}{2} \lambda (\phi).
\end{equation}

 The brane is
located at $r=0$. The r-direction is assumed to be $S^{1}/\Zed_{2}$
instead of an ordinary interval. This explains the factor one-half in
(\ref{branla}); see Appendix for more details.

The equations of motion (\ref{equati1}), (\ref{equati2}) and
(\ref{eqmop}), (\ref{auxeq}) and (\ref{eismatt}), together with the Brane-World ans{\"a}tze
(\ref{m5d}), (\ref{branla}), reduce to:

\begin{eqnarray}
A'' &=& -\frac{2}{3} \phi'^{2} \label{bw1} \\
A'^{2} &=&-\frac{1}{3} V (\phi ) +\frac{1}{6}\phi'^{2} \label{bw2} \\
\frac{\partial V}{\partial \phi } &=& 4 A' \phi' +\phi''\label{bw3}
\end{eqnarray}
where the prime goes for differentiation with respect to $r$.

The junction (or boundary) conditions (\ref{junc}) (or (\ref{extrc0}))
and (\ref{nerrr}) on the brane simplify to (see also \cite{DFGK}): 

\begin{eqnarray}
A'&=&-\frac{1}{3} \lambda \label{junc1}\\
\phi' &=& \frac{1}{2} \frac{\partial \lambda}{\partial \phi }.
\end{eqnarray}

Now, it is straightforward to realize that the expression (\ref{ghhh})
(together with (\ref{that})) {\it vanishes}\footnote{In fact,
(\ref{ghhh}) vanishes for any $\xi^{\mu}$.} with the ansatz
(\ref{branla}):

\begin{equation}\label{enb}
U^{\mu \nu}_{\xi}t_{\mu }n_{\nu} = 0.
\end{equation}




In particular, the energy vanishes.
In previous works \cite{ST,DFGK,Ka} the energy for the simple model
(\ref{m5d}-\ref{branla}) was defined as
minus the Lagrangian (\ref{lagacf})+(\ref{newgr}), {\it without} the
surface term 
(\ref{sdef}). In the static configuration (\ref{m5d}) we
expect to find an 
agreement between the energy and $-\Lg$. 
This is indeed the case when this boundary term $\partial_{\mu}
S^{\mu}$ (\ref{sdef}) is properly taken into account. Using the
equations (\ref{m5d}-\ref{branla}) in (\ref{lagacf})+(\ref{newgr}),
we find:

\begin{eqnarray}
\Hb =-\Lg &=& e^{4A} \left(5 A'^{2}+2 A''+\frac{1}{2}\phi'^{2}+V
(\phi)  \right) -\frac{1}{2}\partial_{r}\left( e^{4A}\lambda\right)\nonumber \\
& & - 2 \partial_{r}\left(e^{4A} A' \right) \label{enhh}
\end{eqnarray}

In the second line of (\ref{enhh}) we isolated the contribution of
$\partial_{\mu} 
S^{\mu}$. In the first line we just recovered the
proposal of \cite{ST,DFGK,Ka} for the energy.

Now using the equations of motion (\ref{bw2}) to replace $V (\phi)$
and then
(\ref{bw1}) to eliminate $\phi '$, it is easy to
check that the 
first term in the rhs of (\ref{enhh}) gives $\partial_{r}
(A'e^{4A})/2$. Then, using the boundary condition (\ref{junc1}),
we find that the total Hamiltonian $\Hb$ indeed vanishes on-shell, in
agreement with 
(\ref{enb}). The vanishing of the energy in the supersymmetric
extension was also pointed out in \cite{BKP}.

The inclusion of the boundary term\footnote{Remember that the boundary
term $\partial_{\mu }S^{\mu }$ is equivalent to the Gibbons and
Hawking term using the ansatz (\ref{nsimpl}).} $\partial_{\mu}
S^{\mu}$ (\ref{sdef}) in the gravitational Lagrangian is quite important since it
ensures that the 
variational principle is satisfied. This criterion allows us to fix the
surface term ambiguity in the Lagrangian 
and cannot be neglected. The gravitational Lagrangian of
a static solution will always be on-shell a boundary term. If this boundary
term is not fixed by some appropriate criterion, as the variational
principle, then the Lagrangian (and so the energy) becomes completely
arbitrary. 

We will finish with another example. 
The ``Minkowski case'' (\ref{m5d}) can be generalized to the de Sitter
one by using 

\begin{equation}\label{desitt}
dS_{4}: h_{ij} dx^{i}dx^{j} = -dt^{2}+e^{2 \sqrt{\bar{\Lambda} } t} (dx^{2}_{1}+dx^{2}_{2}+dx^{2}_{3})
\end{equation}
in the general ansatz (\ref{usform}).

Since the brane Lagrangian is still the simple expression
(\ref{branla}), the equation (\ref{ghhh}) again predicts the vanishing
of the energy (\ref{enb}). Now, this cannot be directly compared to
minus the Lagrangian 
$-\li$ because the metric (\ref{desitt}) is no longer static. In
fact, a direct calculation gives (on-shell)

\begin{eqnarray}
-\li &=& \frac{1}{2} \partial_{r}\left((A'-\lambda (\phi ))
e^{4A+3\sqrt{\bar{\Lambda} } t} \right)-\frac{3}{2}\bar{\Lambda}
e^{4A+3\sqrt{\bar{\Lambda} } t} \nonumber\\
& & +\frac{3}{2} \partial_{t} \left( \sqrt{\bar{\Lambda}} e^{2A+3\sqrt{\bar{\Lambda}
} t} \right) - 2\partial_{r}\left(A'e^{4A+3\sqrt{\bar{\Lambda} } t} \right)\label{tyu}
\end{eqnarray}
where both terms in the second line come from the additional surface
term $\partial_{\mu}S^{\mu}$ (\ref{sdef}).

Again, the total boundary term at $r=0$ vanishes using the junction
condition (\ref{junc1}).
Since the metric (\ref{desitt}) is not static, the total Hamiltonian receives a
non-vanishing contribution from the  $\pi\dot{g}$ term. An explicit
calculation gives:

\begin{eqnarray}\label{enhhg}
\Hb  &=& \pi_{\mu \nu }  \dot{g}^{\mu \nu } -\li\nonumber\\
&=& -3 \bar{\Lambda}
e^{4A+3\sqrt{\bar{\Lambda} } t} - \li=0
\end{eqnarray}
where the canonical momenta is defined as usual by

\begin{equation}\label{defpi}
\pi_{\mu \nu
}:=\frac{\partial \Lg }{\partial \dot{g}^{\mu \nu }}=\frac{\sqrt{\left|
g\right|}}{4 \kappa ^{2}} \left( \tilde{\Kc}_{\mu \nu } -h_{\mu \nu}
\tilde{\Kc} \right),
\end{equation}
with
now $\tilde{\Kc}_{\mu \nu }=\nabla_{(\mu } \hat{t}_{\nu )}$ (with
$\hat{t}_{\mu}=\delta_{\mu}^{0} e^{A}$) the extrinsic curvature on a
Cauchy hypersurface.

Then, the complete Hamiltonian (\ref{enhhg}) indeed vanishes, again in
agreement with our straightforward result (\ref{enb}).
A similar calculation can be repeated for the anti-de Sitter case.

\bigskip

{\bf Acknowledgments.}

I would like to thank R. Dick and B. Julia for enlightening
discussions. I am also grateful to AEI for hospitality.
This work was supported by the EU TMR contract FMRX-CT96-0012.

\section*{Appendix: The Lanczos-Israel equation from the variational principle}

We recall in this appendix the equivalence between
the ``variational principle'' equation (\ref{junc}) and the
Lanczos-Israel junction 
conditions \cite{LI}.
This result, first realized by Hayward and Louko \cite{HL}, can also
be found in 
\cite{CR,DM,Gl,BV}. 

The starting point is equation (\ref{junc}) (together with equation
(\ref{qdef})):

\begin{equation}\label{eww}
\frac{1}{2 \kappa^{2}} \left(n_{\mu } \Gamma^{\mu}_{\rho \sigma} -\frac{1}{2}
g_{\rho \sigma} n_{\mu} \Gamma^{\mu}_{\alpha \beta} g^{\alpha \beta}-\Gamma
_{\alpha (\rho }^{\alpha} n_{\sigma)} +\frac{1}{2}
\Gamma^{\alpha }_{\beta \alpha} n^{\beta} g_{\rho
\sigma}\right) =T^{bra}_{\rho \sigma  }.
\end{equation}

If we contract equation (\ref{eww}) with $g^{\rho \sigma}$ and put it
back into (\ref{eww}), we obtain:

\begin{equation}\label{egh}
\frac{1}{2 \kappa^{2}} \left(n_{\mu } \Gamma^{\mu}_{\rho \sigma}-\Gamma
_{\alpha (\rho }^{\alpha} n_{\sigma)}\right) = T_{\rho
\sigma}^{bra}-\frac{1}{D-2}g_{\rho \sigma } T^{bra}.
\end{equation}

We can project the above equation with $h^{\rho}_{\mu }h^{\sigma
}_{\nu }$ (remember that
$h^{\rho}_{\mu}=\delta^{\rho}_{\mu}-\hat{n}^{\rho} \hat{n}_{\mu}$) and
use the fact that $n_{\mu }=\hat{n}_{\mu}$ is constant\footnote{The
result (\ref{extrc}) can also be recovered by adding the Gibbons and
Hawking \cite{GH} surface term (instead of (\ref{sdef})) to the gravitational
Lagrangian \cite{HL,CR}.} (see equation (\ref{nsimpl})) to rewrite it as:

\begin{equation}\label{extrc}
-\frac{1}{2 \kappa^{2}} \Kc_{\mu \nu} = T_{\mu \nu}^{bra}-\frac{1}{D-2}h_{\mu \nu} T^{bra},
\end{equation}
with $\Kc_{\mu \nu}:=\nabla_{(\mu} n_{\nu)}$ the extrinsic curvature on
the brane.

If the brane is locate in the bulk (not at one
boundary), the same variational principle argument modifies
(\ref{extrc}) to \cite{HL}:

\begin{equation}\label{extrc2}
-\frac{1}{2 \kappa^{2}} \left(   \Kc_{\mu \nu}^{+}-  \Kc_{\mu
\nu}^{-}\right)= T_{\mu \nu}^{bra}-\frac{1}{D-2}h_{\mu \nu} T^{bra},
\end{equation}
which are the usual Lanczos-Israel junction conditions \cite{LI}
(remember that $2\kappa ^{2}=8 \pi G$).

In this manuscript, we worked with one brane located
at the boundary $r =0$. If we consider the orbifold $S^{1}/\Zed_{2}$
instead of an ordinary interval, the equation (\ref{extrc2}) has to be
completed  with
$\Kc_{\mu \nu}^{+}=-  \Kc_{\mu \nu}^{-} =\Kc_{\mu \nu}$. The net
result is that equation (\ref{extrc}) is modified by a factor of
two (see section \ref{exampl} for examples): 

\begin{equation}\label{extrc22}
-\frac{1}{2 \kappa^{2}} \Kc_{\mu \nu} = \frac{1}{2} \left( T_{\mu \nu}^{bra}-\frac{1}{D-2}h_{\mu \nu} T^{bra}\right).
\end{equation}
Note that the rhs of the junction condition (\ref{nerrr}) for the scalar fields
is also modified by a factor $1/2$. 

We can then repeat the calculation of the superpotential using now the
modified boundary conditions (\ref{extrc22}) in order to ``integrate''
equation (\ref{begeq}). It is straightforward to check that the final
superpotential on the orbifold is one half the one on the plain interval:

\begin{equation}\label{halfo}
\ _{(S^{1}/\tiny{\Zed_{2}})} U^{\mu \nu }_{\xi }t_{\mu }n_{\nu } = \frac{1}{2}
 U^{\mu \nu }_{\xi }t_{\mu }n_{\nu }.
\end{equation}

The simplest way to take into account this $1/2$ factor which appears
in the orbifold case is to make the replacement
$\Lg_{bra}\rightarrow \Lg_{bra}/2$ 
(and then $T^{bra}_{\rho \sigma} 
\rightarrow T^{bra}_{\rho \sigma}/2$) in all the above formulas.
In other words, the orbifold case can be treated using the plain
segment equations together with a rescalling of the brane Lagrangian
by one half.

We would like to emphasize that this factor of one half is given by the
manifold considered. It is
therefore an extra imput to be adjusted according to the physical
problem studied.

\end{document}